%% file: main.tex
\documentclass[a4paper,amsfonts,amssymb,amsmath, reprint,showkeys,superscriptaddress,nofootinbib, twoside]{revtex4-2}
\usepackage[english]{babel}
\usepackage[utf8]{inputenc}
\usepackage[colorinlistoftodos, color=green!40, prependcaption]{todonotes}
\input{preamble}

\usepackage[pdftex, pdftitle={Article}, pdfauthor={Author}]{hyperref}
 
\begin{document}

\title{Rapid on-demand generation of thermal states in superconducting quantum circuits}
 
\author{Timm Mörstedt}
\email[Corresponding author.\\]{timm.morstedt@aalto.fi}
\affiliation{QCD Labs, QTF Centre of Excellence, Department of Applied Physics, Aalto University, P.O. Box 13500, FI-00076 Aalto, Finland}
\author{Wallace S. Teixeira}
\affiliation{QCD Labs, QTF Centre of Excellence, Department of Applied Physics, Aalto University, P.O. Box 13500, FI-00076 Aalto, Finland}
\author{Arto Viitanen}
\affiliation{QCD Labs, QTF Centre of Excellence, Department of Applied Physics, Aalto University, P.O. Box 13500, FI-00076 Aalto, Finland}
\author{Heidi Kivijärvi}
\affiliation{QCD Labs, QTF Centre of Excellence, Department of Applied Physics, Aalto University, P.O. Box 13500, FI-00076 Aalto, Finland}
\author{Maaria Tiiri}
\affiliation{QCD Labs, QTF Centre of Excellence, Department of Applied Physics, Aalto University, P.O. Box 13500, FI-00076 Aalto, Finland}
\author{Miika Rasola}
\affiliation{QCD Labs, QTF Centre of Excellence, Department of Applied Physics, Aalto University, P.O. Box 13500, FI-00076 Aalto, Finland}
\author{Andras Marton Gunyho}
\affiliation{QCD Labs, QTF Centre of Excellence, Department of Applied Physics, Aalto University, P.O. Box 13500, FI-00076 Aalto, Finland}
\author{Suman Kundu}
\affiliation{QCD Labs, QTF Centre of Excellence, Department of Applied Physics, Aalto University, P.O. Box 13500, FI-00076 Aalto, Finland}
\author{Louis Lattier}
\affiliation{QCD Labs, QTF Centre of Excellence, Department of Applied Physics, Aalto University, P.O. Box 13500, FI-00076 Aalto, Finland}
\author{Vasilii Vadimov}
\affiliation{QCD Labs, QTF Centre of Excellence, Department of Applied Physics, Aalto University, P.O. Box 13500, FI-00076 Aalto, Finland}

\author{Gianluigi Catelani}
\affiliation{JARA Institute for Quantum Information (PGI-11), Forschungszentrum Jülich, 52425 Jülich, Germany}
\affiliation{Quantum Research Center, Technology Innovation Institute, Abu Dhabi 9639, UAE}
\author{Vasilii Sevriuk}
\affiliation{IQM Quantum Computers, FI-02150 Espoo, Finland}
\author{Johannes Heinsoo}
\affiliation{IQM Quantum Computers, FI-02150 Espoo, Finland}
\author{Jukka Räbinä}
\affiliation{IQM Quantum Computers, FI-02150 Espoo, Finland}
\author{Joachim Ankerhold}
\affiliation{Institut für Komplexe Quantensysteme and IQST, Universität Ulm, Ulm, Germany}
\author{Mikko Möttönen}
\email[Corresponding author.\\]{mikko.mottonen@aalto.fi}
\affiliation{QCD Labs, QTF Centre of Excellence, Department of Applied Physics, Aalto University, P.O. Box 13500, FI-00076 Aalto, Finland}
\affiliation{QTF Centre of Excellence, VTT Technical Research Centre of Finland Ltd., P.O. Box 1000, 02044 VTT, Finland}

\begin{abstract} 


We experimentally demonstrate the fast generation of thermal states of a transmon using a single-junction quantum-circuit refrigerator (QCR) as an in-situ-tunable environment. Through single-shot readout, we monitor the transmon up to its third-excited state, assessing population distributions controlled by QCR drive pulses. Whereas cooling can be achieved in the weak-drive regime, high-amplitude pulses can generate Boltzmann-distributed populations from a temperature of 110~mK up to 500~mK within 100~ns. As we propose in our work, this fast and efficient temperature control provides an appealing opportunity to demonstrate a quantum heat engine. Our results also pave the way for efficient dissipative state preparation and for reducing the circuit depth in thermally assisted quantum algorithms and quantum annealing.

\end{abstract}

\include{math-macros-v2}

\include{symb-macros-v3}

\maketitle

\textit{Introduction.}--With superconducting qubits emerging as building blocks of a potential quantum computer, fast and precise control over a large number of qubits is a prerequisite for reliable execution of quantum algorithms ~\cite{DiVincenco2000,Manenti2021}. In particular, the high-fidelity preparation of initial quantum states for these algorithms strongly affects their accuracy. Although this has been widely studied and optimized for qubit eigenstates and superpositions thereof~\cite{Yang2019}, quickly reaching the equilibrium state of a given temperature, the Gibbs state, while keeping the ambient temperature low appears challenging~\cite{Sagastizabal2021, Wang2021}.

Beyond specific state preparation, active control over thermal excitations in superconducting circuits can be an important building block for quantum thermodynamics~\cite{Pekola2015} in general, such as for quantum-heat-transport experiments~\cite{Ronzani2018} and quantum heat engines~\cite{Alicki1979}. Accurately prepared thermal states~\cite{Hsu2023} can also be used in the context of minimal heat engines where a qubit is coupled to bandgap reservoirs~\cite{Xu2022a} which may also serve as a starting point to explore qubit-environment correlations~\cite{Xu2022b,Tuorila2019}. Local cooling of different types of superconducting circuits has already been demonstrated with quantum-circuit refrigerators (QCR)~\cite{Tan2017,Silveri2017,Sevriuk2019,Sevriuk2022,Viitanen2023} which are a small on-chip dissipator that can be exponentially turned on and off using a bias voltage. Interestingly, a QCR accurately represents a thermal environment for the electromagnetic degrees of freedom and can correspondingly heat a quantum circuit at high bias voltages, offering a versatile two-way tunability: cooling or heating as desired. In particular, a qubit or resonator can be heated or cooled with the QCR in thermodynamic cycles without the presence of other thermal baths. As an example, in a quantum Otto cycle, the QCR may be activated to heat or cool the system during the isochoric process, and it would remain off during adiabatic excursions~\cite{Teixeira2022}. Furthermore, it can be used to charge a fixed-heat-capacity element, or a heat battery. The charged heat may then be used to fuel thermal devices or experiments. Finally, a modified QCR comprising a quantum dot has been proposed as a possible cryogenic source of microwaves~\cite{Hsu2021} 


\begin{figure*}
    \subfloat{\label{fig:setupa}}
    \subfloat{\label{fig:setupb}}
    \subfloat{\label{fig:setupc}}
    \subfloat{\label{fig:setupd}}
    \includegraphics[width=1.0\linewidth]{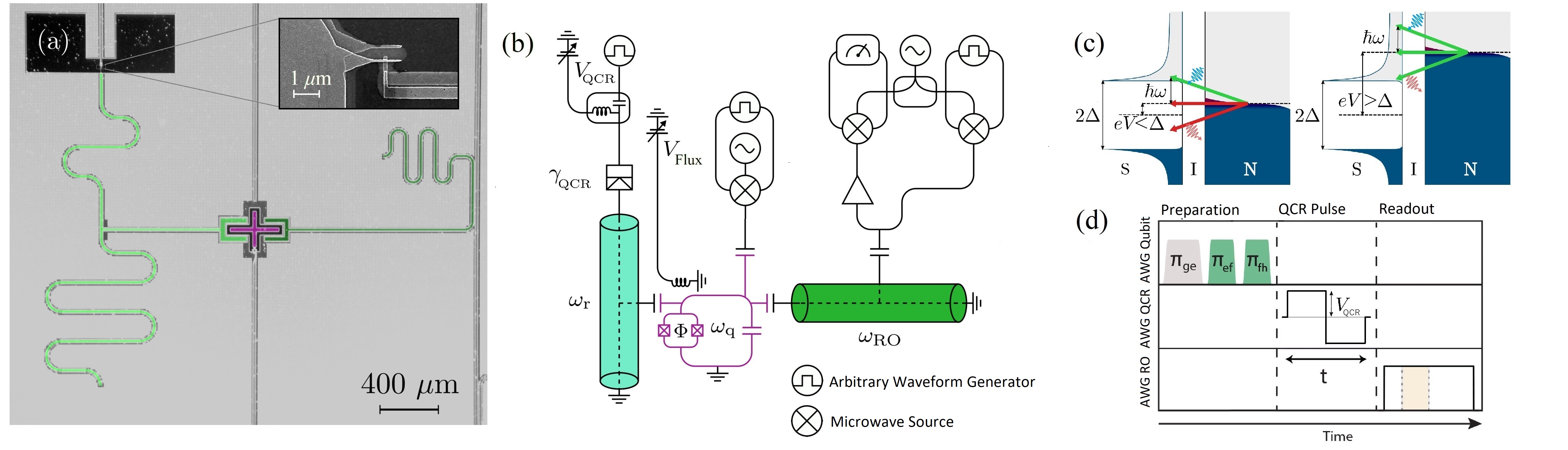} 
    \caption{Measurement setup and device. (a) False-color optical micrograph of the sample. The transmon qubit (purple) is coupled to the reset resonator (light green) which is connected to the input line through the QCR (inset). The QCR is oriented such that the normal-conductor side of the NIS junction connects to the input line, whereas the superconducting Al part is in galvanic contact with the Nb resonator. The qubit is coupled to a separate resonator for readout (dark green). (b)~Simplified circuit diagram of the chip and the experimental setup with colors matching to those in (a). The transmon qubit is represented by two parallel Josephson junctions and a capacitor, and it is controlled through a dc flux line of the junction loop and an rf drive line for qubit state preparation. The QCR is shown as an NIS junction biased with a dc source for bias offset and an arbitrary waveform generator (AWG) for pulsing. (c)~Tunneling process in the NIS junction depending on bias voltage $V$. If $|eV|<\Delta$ (left), tunneling is only possible with the absorption of a photon (cooling), whereas higher voltages (right) also allow elastic tunneling and photon emission, leading to heating. (d)~Pulse sequence: Three different arbitrary waveform generators (AWGs) are used to create subsequent state preparation, QCR, and readout pulses. Preparation pulses are only required for the calibration measurement in Fig.~\ref{fig:result}(a), and are not applied in heating experiments. The QCR pulse is a net-zero square pulse with a fundamental frequency of 100~MHz.
    }
    \label{fig:setup}
\end{figure*}

\textit{Sample and its operation Principle.}--To realize an on-demand thermal state in a superconducting qubit, we carry out experiments with the sample depicted in Fig.~\ref{fig:setup}(a). Here, a transmon~\cite{Koch2007}, with qubit frequency $\omega_{\text{ge}}/(2\pi)=4.09$ GHz and anharmonicity $\alpha/(2\pi)=-273$ MHz, is coupled to a quantum-circuit refrigerator (QCR) via a coplanar-waveguide (CPW) resonator of frequency $\omega_{1}/(2\pi)=4.67$ GHz.  
In addition, a CPW resonator of frequency $\omega_{2}/(2\pi)=7.44$ GHz is coupled to the transmon for the standard dispersive readout of the state of the transmon~\cite{Blais2004}. The core Hamiltonian including the coupling of the transmon to both resonators can be expressed as
\begin{align}
    \hH/\hbar=&\ \omega_{\text{ge}}\hdgg{b} \hb+\frac{\alpha}{2}\hdgg{b}\hdgg{b}\hb\hb+\omega_{1}\hdgg{a}_1 \ha_{1} + \omega_{2}\hdgg{a}_{2}\ha_{2}\nonumber\\
    {}&+g_1\left(\hdgg{b}\ha_{1}+\hb\hdgg{a}_{1}\right)+g_2\left(\hdgg{b}\ha_{2}+\hb\hdgg{a}_{2}\right),\label{eq:H}
\end{align}
where $\hb$ and $\ha_j$ ($j=1,2$) are the annihilation operators of the transmon and the resonator $j$, respectively. The relatively weak coupling frequencies, $g_1/(2\pi)=59.6$~MHz and $g_2/(2\pi)=70.4$~MHz, place the transmon in the dispersive regime with both resonators~\cite{Blais2021}. Consequently, this geometry allows for strong dissipation control without compromising the qubit lifetime in the QCR-off state~\cite{Sevriuk2022} and for advanced qubit reset protocols~\cite{Yoshioka2021, Yoshioka2023,Teixeira2024}.  

The QCR consists of a normal-metal--insulator--superconductor (NIS) junction that can promote either heating or cooling of the core circuit depending on its bias voltage, $V_{\text{QCR}}$. The underlying mechanism is the quasi-particle tunneling through the junction illustrated in Fig.~\ref{fig:setup}(c)~\cite{Silveri2017, Hsu2020, Vadimov2022}. If $|V_{\text{QCR}}|< \Delta/e$, where $e$ is the elementary charge and $\Delta=0.215$~meV is the superconductor gap parameter of aluminum, quasi-particle tunneling is
ideally only possible with a simultaneous absorption of energy exceeding $\Delta-e|V_{\text{QCR}}|$, thus cooling the circuit. A part of the energy may be given by the thermal excitations of the quasiparticles in the normal metal, leading to an exponential rise of the QCR-induced decay rate with the bias voltage. However, bias voltages $|V_{\text{QCR}}|> \Delta/e$ induce photon emission in the course of the tunneling, leading to heating of the core circuit. Thus, controlling $V_{\text{QCR}}$, we may modify both decay rates and the thermal occupation number of the core circuit. Through single-shot dispersive readout, in this work, we show that transmon temperatures up to $500$~mK can be achieved in a time-window of $100$~ns, demonstrating a remarkable progress towards applications with controlled thermal states.

\textit{Sample Fabrication and Experimental Setup.}--The superconducting circuit shown in Fig.~\ref{fig:setup}(a) is fabricated by optical lithography and reactive ion etching of a 200-nm film of sputtered niobium on a six-inch intrinsic-silicon wafer. The Josephson junctions are added in a separate step through electron beam lithography (EBL), creating an undercut in a double-layer resist of a methyl methacrylate copolymer
and poly(methyl methacrylate). Aluminum is deposited using a Dolan-bridge two-angle shadow evaporation technique with in-situ oxidation to form the tunnel barrier in between two Al layers. Similarly to the Josephson junctions, the NIS junction is fabricated in EBL, but replacing Al with 60~nm of Cu in the top layer. In addition, a 3-nm layer of Al is deposited as an adhesion layer between aluminum oxide and copper. Due to the inverse proximity effect, this adhesion layer is not superconducting, but part of the normal-conducting electrode of the NIS junction. The adhesion layer reduces oxidation and diffusion of the Cu layer, significantly improving the yield and longevity of the device.

\begin{figure}
    \subfloat{\label{fig:characterizationa}}
    \subfloat{\label{fig:characterizationb}}
    \subfloat{\label{fig:characterizationc}}
    \subfloat{\label{fig:characterizationd}}
    \includegraphics[width=1.0\linewidth]{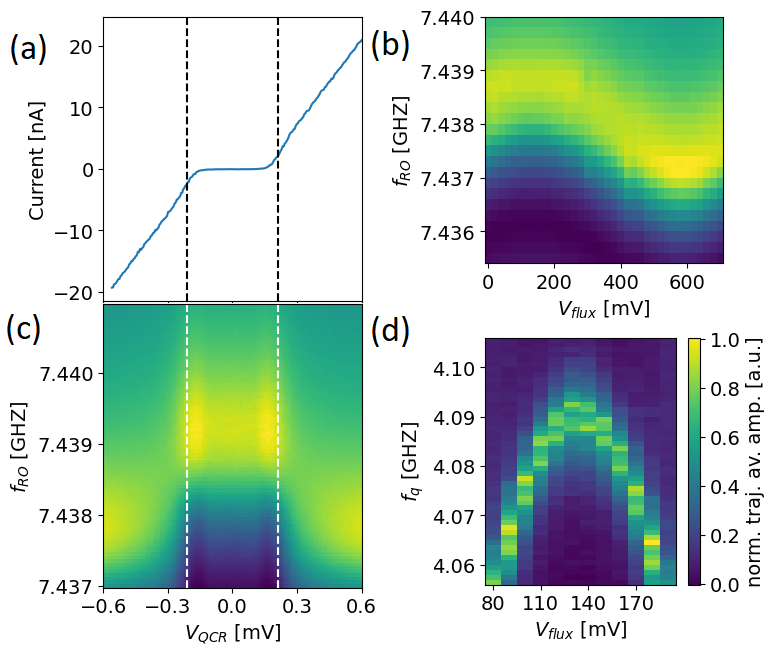} 
    \caption{Resonator, qubit, and QCR characterization. (a) Current through the QCR as a function of voltage across it, providing the superconductor gap and Dynes parameters. (b) Normalized trajectory average amplitude of the tone channeled through the qubit readout resonator as a function of voltage that induces flux through junction loop of the qubit and the frequency of the tone. (c) Trajectory average amplitude of the a tone channeled through the qubit readout resonator as a function of the QCR dc voltage $V_{QCR}$ applied at the junction and the frequency $f_{RO}$ of the tone. The frequency shift of 1.5~MHz occurs at the gap voltage obtained form (a). (d) Average qubit readout signal as a function of the flux-inducing voltage $V_{flux}$ and the frequency of a qubit excitation tone $f_q$ around the flux sweet spot at 130~mV with the QCR turned off. All following measurements are conducted at the sweet spot.
    }
    \label{fig:characterization}
\end{figure}

All measurements shown below are carried out in a dilution refrigerator at a mixing-chamber temperature of 30~mK utilizing a circuit depicted in Fig.~\ref{fig:setup}(b). The QCR control consists of two inputs: a dc component to apply a zero-bias offset determined through an current-voltage (IV) measurement  in Fig.~\ref{fig:characterization}(a) through a dc line filtered with Thermocoax wiring, and a net-zero square pulse applied through an rf coaxial cable. The two signals are combined in a bias tee before reaching the NIS junction. In addition, the qubit can be driven with a weakly coupled rf drive line and read through a separate quarter-wavelength CPW resonator coupled to a transmission line. Current-voltage measurements of the QCR are recorded throught the dc line with a source-measure unit and the qubit readout is carried out as single shots. A readout pulse is applied with a duration of 2~\textmu{s}, but integrated only in the interval 400--800~ns to maximize trajectory separation between different states in the in-phase--quadrature-phase (IQ) plane~\cite{Krantz2019, Walter2017}.

\begin{table}[]
\caption{Measured sample parameters at 30~mK except for the QCR normal-state resistance measured at room temperature.}
\label{parameters}
\centering
\begin{tabular}{|c|c|c|}
\hline
Parameter & Symbol & Value\\
\hline
Qubit frequency & $\omega _\mathrm{ge}/2\pi$ & $4.09\,\mathrm{GHz}$\\
Qubit anharmonicity & $\alpha/2\pi$ & $-273\,\mathrm{MHz}$\\
Reset resonator frequency & $\omega _\mathrm{1}/2\pi$ & $4.67\,\mathrm{GHz}$\\
Readout resonator frequency & $\omega _\mathrm{2}/2\pi$ & $7.44\,\mathrm{GHz}$\\
Reset coupling strength & $g_\mathrm{1}/2\pi$ & $59.6\,\mathrm{MHz}$\\
Readout coupling strength & $g_\mathrm{2}/2\pi$ & $70.4\,\mathrm{MHz}$\\
QCR resistance & $R_\mathrm{QCR}$ & $13.8\,\mathrm{k \Omega}$\\
Dynes parameter & $\gamma_\mathrm{D}$ & $2.3\times 10^{-3}$\\
\hline
\end{tabular}
\end{table}

\begin{figure}
    \subfloat{\label{fig:resulta}}
    \subfloat{\label{fig:resultb}}
    \subfloat{\label{fig:resultc}}
    \subfloat{\label{fig:resultd}}
    \includegraphics[width=1.0\linewidth]{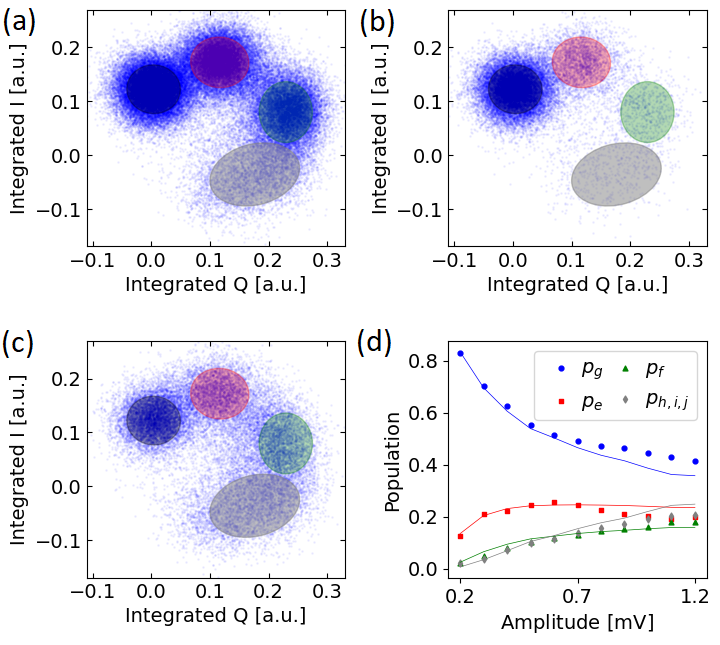} 
    \caption{Single-shot measurements with 10000 shots per prepared qubit state. (a)~Calibration measurement of all prepared states g--h superimposed and the corresponding fit to a Gaussian-mixture model with four components. The shaded ellipses for states g (blue), e (red), f (green), and h (gray) represent the $1\sigma$ covariances of each qubit state. The h-state fit possibly includes even higher-energy states due to its large width. (b)~As (a) but for the thermal equilibrium state at $V_\textrm{QCR}=0$ without any state preparation pulses. (c)~As (b), but after a 100-ns QCR pulse with amplitude $V_\textrm{QCR}=1.2$~mV. The population shift toward high-energy states arises from heating caused by the high pulse amplitude $V_\textrm{QCR}>\Delta$. (d)~Qubit state populations extracted from single-shot measurements as functions of the QCR pulse amplitude for a fixed pulse length of $100\,\mathrm{ns}$, compared to a fitted Boltzmann distribution truncated at the sixth transmon state.
    Owing to the wide distribution of the h-state, we assume that the states i and j are mostly contained within the $1\sigma$ boundary of the h-state. }
    \label{fig:result}
\end{figure}

\textit{Results.}--The experimental results to characterize the key sample parameters are shown in Fig.~\ref{fig:characterization} and the extracted parameters are listed in Tab.~\ref{parameters}. From the IV curve of the QCR junction in Fig.~\ref{fig:characterization}(a) we extract the superconductor gap parameter $\Delta=0.215$~meV as the halfwidth of the plateau as well the Dynes parameter $\gamma_\mathrm{D}$. Assuming $\gamma_\mathrm{D}\ll1$, the latter can be approximated as the ratio of resistances in and outside of the gap voltage region, yielding $\gamma_\mathrm{D}=2.3\times 10^{-3}$.

Furthermore, Fig.~\ref{fig:characterization}(c) shows a 1.5-MHz resonance shift in the readout resonator caused by the QCR bias at the gap voltage obtained from Fig.~\ref{fig:characterization}(a). While the QCR amplitude creates a Lamb shift in the reset resonator, this causes a dispersive shift in the coupled circuit, which we can measure at the readout resonator, together with the Lamb shift directly induced by the QCR on the readout resonator~\cite{Viitanen2021}. For all following experiments, the applied pulse sequence is designed such that any QCR pulse is turned off before the beginning of the readout pulse, and hence this shift is absent during qubit readout. 

Qubit measurements are carried out in single shots and shown in Fig.~\ref{fig:result}. For calibration measurements, we prepare each of the four lowest-energy states of the transmon through a corresponding sequence of $\pi$ pulses, as shown in Fig.~\ref{fig:setup}(d). Combining the single-shot readout data from all these initial states into a single distribution in the IQ plane, four prominent point clouds are obtained. We use a multivariate Gaussian mixture model (GMM) to fit the combined single-shot calibration data as shown in Fig.~\ref{fig:result}(a). The ratio of points in- and outside of a given Mahalanobis distance $m\sigma$ from each state for an $n$-dimensional Gaussian distribution is given by the generalized regularized incomplete gamma function $Q(n/2, 0, (m^2)/2)$. Choosing the $1\sigma$ ellipsoid of a two-dimensional Gaussian, we find that $Q(1, 0, 0.5)=39.34\%$ of the measured points of each state should be within its $1\sigma$ boundary~\cite{Bajorski2012}. The count of shots inside these boundaries can be used to extract the corresponding qubit state populations, allowing us to normalize the population truncated at the fourth state. In addition, we include a correction matrix to account for the overlap of the distributions of different states in the IQ plane. The Gaussian fit of the h-state distribution is by far the least precise due to reduced state preparation fidelity, its width, and proximity to states that are not included in the GMM. Given the overestimation of the h-state population throughout the experiment, we assume that the h-state fit includes measurements of the i-state and the j-state. 

Figures~\ref{fig:result}(b) and~\ref{fig:result}(c) highlight the difference of the single-shot distributions with and without the QCR pulse on the thermal-equilibrium state. In the QCR-off state, the thermal equilibrium population of the g-state is $83\%$, which is lower than a typical transmon at low temperature~\cite{Jin2015}, but $8\%$ higher than in previous experiments of a similar circuit with a QCR~\cite{Yoshioka2021}. After a 100-ns-long QCR pulse of 1.2-mV amplitude, the ground-state population has been reduced to 42$\%$, half of its original value, suggesting a significant rise in temperature. If the initial state is non-thermal, for example e or f, the resulting state after the QCR pulse is also non-thermal, but approaches a thermal state for much longer QCR pulses than the qubit $T_1$ at the given QCR bias.

\begin{figure}
    \subfloat{\label{fig:discussiona}}
    \subfloat{\label{fig:discussionb}}
    \includegraphics[width=1.0\linewidth]{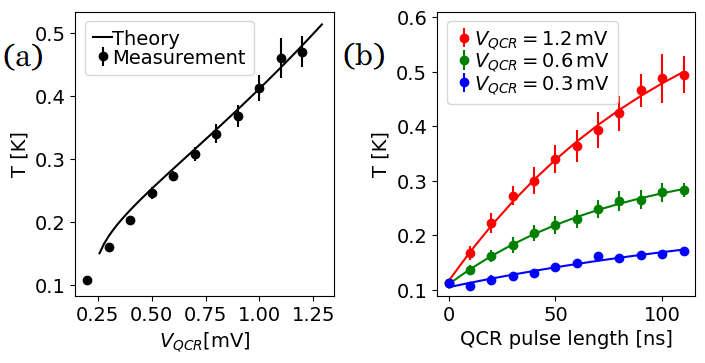} 
    \caption{Temperature $T$ of the closest Gibbs state extracted from single shot measurements of the transmon. (a) Transmon temperature as function of QCR pulse amplitude for a fixed pulse time of 100~ns (markers) agrees well with the theory model for energies above the superconducting gap (solid line) as expected from Ref.~\cite{Hsu2020}, see supplemental material (SM) for details. For large pulse amplitudes, the heating rate of 0.36~K/mV agrees well with the measured data well above the gap voltage $\Delta/e$. The errorbars represent the deviation of the measured states from an ideal Gibbs state, for example due to truncation errors. (b) Measured qubit temperature (markers) as a function of the QCR pulse length for three different amplitudes above the gap voltage as indicated. We also show fitted exponential functions of the form $T(t)=T_0+A(1-e^{-\frac{t}{\tau}})$ (solid lines) yielding a time constant $\tau=185\,\mathrm{ns}$ for $V_{\mathrm{QCR}}=0.3\,\mathrm{mV}$, $\tau=80\,\mathrm{ns}$ for $V_{\mathrm{QCR}}=0.6\,\mathrm{mV}$, and $\tau=109\,\mathrm{ns}$ for $V_{\mathrm{QCR}}=1.2\,\mathrm{mV}$.
    }
    \label{fig:discussion}
\end{figure}

 Figure~\ref{fig:result}(d) shows the populations of the four lowest-energy states of the transmon after a 100-ns QCR pulse of different amplitudes together with least-square fits to Gibbs states. We observe that the data follows resonably well the Gibbs state with the temperature increasing for increasing amplitudes. Figure~\ref{fig:discussion}(a) shows the resulting temperatures, from which we conclude that we may choose at will an amplitude to prepare the state of the transmon to any temperature from 110~mK to 476~mK with the 100-ns pulse. The errorbars for temperature estimates are given by the fitting covariance and therefore represent the deviation from an ideal Gibbs state. This deviation is significantly large for high temperatures: it remains below 2$\%$ for temperatures up to 200~mK, climbing to 19$\%$ at $T=470\,\mathrm{mK}$. This increase in uncertainty can be attributed to truncation errors, as only the four lowest-energy states are measured with normalized populations. In addition, a large population of higher states further aggravates the overestimation of the h-state. Importantly, quantum fidelities between the transmon state and an ideal thermal state surpassing $99\%$ are theoretically predicted for this sample, see the Supplemental Material for details.
 
 Figure~\ref{fig:discussion}(b) shows the extracted transmon temperature as a function of the pulse length for a different QCR amplitudes. We observe a trend of increasing temperature consistent with Fig.~\ref{fig:discussion}(a). The short-time evolution of the temperature accurately follows an exponential fit of the form $T(t)=T_0+A(1-e^{-\frac{t}{\tau}})$ with $T_0=110$~mK. Although the temperature exhibits some saturation in time, yet it shows a strong increase for increasing amplitudes up to 1.2~mV, suggesting that even higher temperatures are possible within this timescale. Compared to ohmic heating of a normal conductor on the chip, the QCR requires several orders of magnitudes lower power due to the strong coupling of the heating power to the qubit in the QCR-on state and low amount of waste heat owing to the fact that there is no quasiparticle reservoir that needs to be heated up, thus allowing for low ambient temperature although providing high qubit temperatures. Furthermore, the coupled circuit is protected from decoherence in the QCR-off state with weak dissipation and detuning of the reset resonator.

\textit{Discussion and conclusions.}--Different heating mechanisms of the qubit may play a role in this device: Quasiparticles generated at the tunnel barrier can contribute to heating effects even in the QCR-off state, limiting the QCR on-off ratio~\cite{Catelani_2022, Mannila2022, Karzig2021}. Breaking and recombination of Cooper pairs at the QCR junction can also eject ballistic phonons into the  substrate and eventually excite the qubit~\cite{Eisenmenger1976, Maisi2013}. However, the dominant effect in the high-amplitude case of Fig.~\ref{fig:discussion} is photon-assisted tunneling as described in Fig~\ref{fig:setup}(c). This is consistent with the dependence of the temperature on the QCR pulse amplitude in Fig.~\ref{fig:discussion}(a) and with previous high-bias measurements of the temperature of a resonator coupled to a QCR~\cite{Masuda2018}.

Let us next elaborate on the realization of a single-bath quantum heat engine with our QCR device shown in Fig.~\ref{fig:setup}(a). 
Even though a heat engine utilizing only a single static heat reservoir is thermodynamically impossible~\cite{Landi2021}, the QCR in combination with the reset resonator can replace both the cold and hot reservoir of a typical heat engine since it is in-situ tunable. For example, a quantum Otto cycle can be implemented through alternating pulses in the QCR and the flux line as follows: (i) isochoric heating with the QCR at high amplitude ($|V_\textrm{QCR}|>\Delta/e$), (ii) adiabatic lowering of $\omega_\textrm{ge}$ through a flux sweep to $\omega_\textrm{ge}^\textrm{min}$, (iii) isochoric cooling of the circuit with a low-amplitude QCR pulse ($|V_\textrm{QCR}|<\Delta/e$) and (iv) adiabatic tuning of $\omega_\textrm{ge}$ back to $\omega_\textrm{ge}^\textrm{max}$ at the sweet spot~\cite{Kosloff2017}. The pulses for heating and frequency tuning can each be applied on a timescale of less than 100~ns, still shorter than the 350-ns equilibration time of the reset resonator during sweeps (ii) and (iv) but longer than during the thermalization sweeps (i) and (iii) when the QCR is on, thus minimizing its disturbance to the Otto cycle~\cite{Camati2019}. The efficiency of such an Otto heat engine is bounded by the Carnot efficiency $\eta_\textrm{c}=1-\frac{T_\textrm{c}}{T_\textrm{h}}$, where $T_\textrm{c}$ and $T_\textrm{h}$ are the electromagnetic temperatures of the QCR in its cooling and heating mode, respectively. Since QCR provides a large temperature tuning ratio, one may expect efficiencies close to that given by the frequency tuning $\eta_\textrm{f}=1-\frac{\omega_\textrm{ge}^\textrm{min}}{\omega_\textrm{ge}^\textrm{max}}$.

In summary, we have experimentally realized the controlled generation of thermal states up to 0.5~K in a transmon qubit on timescales below 100~ns. Four-state single-shot analysis of the transmon is in good agreement with the expected Gibbs distribution of a thermal state. Unlike simple ohmic heating, we can heat the qubit without a significant increase in ambient temperature thanks to the utilization of a quantum-circuit refrigerator as an in-situ-tunable thermal environment. With its cooling capability already analyzed in previous work~\cite{Tan2017, Sevriuk2022, Viitanen2023}, we have here shown two-way tunability of the environment for superconducting circuits. Our results pave the way for implementations of quantum heat engines and other open quantum systems in superconducting circuits.

This work was funded by the Academy of Finland Centre of Excellence program (project Nos. 352925, and 336810) and grant Nos.~316619 and 349594 (THEPOW). We also acknowledge funding from the European Research Council under Advanced Grant No.~101053801 (ConceptQ) and the provision of facilities and technical support by Aalto University at OtaNano—Micronova Nanofabrication Centre. We thank Jukka Pekola, Bayan Karimi, Jian Ma, Satrya Christoforus, Visa Vesterinen, Tapio Ala-Nissila, and Matti Silveri for discussions. We also thank Jukka-Pekka Kaikkonen and Kimmo Sten at VTT Technical Research Center for fabricating the Nb structures.

\bibliography{refs}

\end{document}

%% file: preamble.tex
\usepackage{amsthm}
\usepackage{mathtools}
\usepackage{physics}
\usepackage{xcolor}
\usepackage{graphicx}
\usepackage{adjustbox}
\usepackage{placeins}
\usepackage[T1]{fontenc}
\usepackage{lipsum}
\usepackage{csquotes}

\usepackage[subrefformat=parens,labelformat=parens,caption=false]{subfig}
\usepackage{collref}

%% file: math-macros-v2.tex

\global\long\def\ket#1{|#1\rangle}%

\global\long\def\Ket#1{\left|#1\right>}%

\global\long\def\bra#1{\langle#1|}%

\global\long\def\Bra#1{\left<#1\right|}%

\global\long\def\bk#1#2{\langle#1|#2\rangle}%

\global\long\def\BK#1#2{\left\langle #1\middle|#2\right\rangle }%

\global\long\def\kb#1#2{\ket{#1}\!\bra{#2}}%

\global\long\def\KB#1#2{\Ket{#1}\!\Bra{#2}}%

\global\long\def\mel#1#2#3{\bra{#1}#2\ket{#3}}%

\global\long\def\MEL#1#2#3{\Bra{#1}#2\Ket{#3}}%

\global\long\def\n#1{|#1|}%

\global\long\def\N#1{\left|#1\right|}%

\global\long\def\ns#1{|#1|^{2}}%

\global\long\def\NS#1{\left|#1\right|^{2}}%

\global\long\def\nn#1{\lVert#1\rVert}%

\global\long\def\NN#1{\left\lVert #1\right\rVert }%

\global\long\def\nns#1{\lVert#1\rVert^{2}}%

\global\long\def\NNS#1{\left\lVert #1\right\rVert ^{2}}%

\global\long\def\ev#1{\langle#1\rangle}%

\global\long\def\EV#1{\left\langle #1\right\rangle }%

%% file: symb-macros-v3.tex
\global\long\def\ha{\hat{a}}%

\global\long\def\hb{\hat{b}}%

\global\long\def\hc{\hat{c}}%

\global\long\def\hd{\hat{d}}%

\global\long\def\he{\hat{e}}%

\global\long\def\hf{\hat{f}}%

\global\long\def\hg{\hat{g}}%

\global\long\def\hh{\hat{h}}%

\global\long\def\hi{\hat{i}}%

\global\long\def\hj{\hat{j}}%

\global\long\def\hk{\hat{k}}%

\global\long\def\hl{\hat{l}}%

\global\long\def\hm{\hat{m}}%

\global\long\def\hn{\hat{n}}%

\global\long\def\ho{\hat{o}}%

\global\long\def\hp{\hat{p}}%

\global\long\def\hq{\hat{q}}%

\global\long\def\hr{\hat{r}}%

\global\long\def\hs{\hat{s}}%

\global\long\def\hu{\hat{u}}%

\global\long\def\hv{\hat{v}}%

\global\long\def\hw{\hat{w}}%

\global\long\def\hx{\hat{x}}%

\global\long\def\hy{\hat{y}}%

\global\long\def\hz{\hat{z}}%

\global\long\def\hA{\hat{A}}%

\global\long\def\hB{\hat{B}}%

\global\long\def\hC{\hat{C}}%

\global\long\def\hD{\hat{D}}%

\global\long\def\hE{\hat{E}}%

\global\long\def\hF{\hat{F}}%

\global\long\def\hG{\hat{G}}%

\global\long\def\hH{\hat{H}}%

\global\long\def\hI{\hat{I}}%

\global\long\def\hJ{\hat{J}}%

\global\long\def\hK{\hat{K}}%

\global\long\def\hL{\hat{L}}%

\global\long\def\hM{\hat{M}}%

\global\long\def\hN{\hat{N}}%

\global\long\def\hO{\hat{O}}%

\global\long\def\hP{\hat{P}}%

\global\long\def\hQ{\hat{Q}}%

\global\long\def\hR{\hat{R}}%

\global\long\def\hS{\hat{S}}%

\global\long\def\hT{\hat{T}}%

\global\long\def\hU{\hat{U}}%

\global\long\def\hV{\hat{V}}%

\global\long\def\hW{\hat{W}}%

\global\long\def\hX{\hat{X}}%

\global\long\def\hY{\hat{Y}}%

\global\long\def\hZ{\hat{Z}}%

\global\long\def\hap{\hat{\alpha}}%

\global\long\def\hbt{\hat{\beta}}%

\global\long\def\hgm{\hat{\gamma}}%

\global\long\def\hGm{\hat{\Gamma}}%

\global\long\def\hdt{\hat{\delta}}%

\global\long\def\hDt{\hat{\Delta}}%

\global\long\def\hep{\hat{\epsilon}}%

\global\long\def\hvep{\hat{\varepsilon}}%

\global\long\def\hzt{\hat{\zeta}}%

\global\long\def\het{\hat{\eta}}%

\global\long\def\hth{\hat{\theta}}%

\global\long\def\hvth{\hat{\vartheta}}%

\global\long\def\hTh{\hat{\Theta}}%

\global\long\def\hio{\hat{\iota}}%

\global\long\def\hkp{\hat{\kappa}}%

\global\long\def\hld{\hat{\lambda}}%

\global\long\def\hLd{\hat{\Lambda}}%

\global\long\def\hmu{\hat{\mu}}%

\global\long\def\hnu{\hat{\nu}}%

\global\long\def\hxi{\hat{\xi}}%

\global\long\def\hXi{\hat{\Xi}}%

\global\long\def\hpi{\hat{\pi}}%

\global\long\def\hPi{\hat{\Pi}}%

\global\long\def\hrh{\hat{\rho}}%

\global\long\def\hvrh{\hat{\varrho}}%

\global\long\def\hsg{\hat{\sigma}}%

\global\long\def\hSg{\hat{\Sigma}}%

\global\long\def\hta{\hat{\tau}}%

\global\long\def\hup{\hat{\upsilon}}%

\global\long\def\hUp{\hat{\Upsilon}}%

\global\long\def\hph{\hat{\phi}}%

\global\long\def\hvph{\hat{\varphi}}%

\global\long\def\hPh{\hat{\Phi}}%

\global\long\def\hch{\hat{\chi}}%

\global\long\def\hps{\hat{\psi}}%

\global\long\def\hPs{\hat{\Psi}}%

\global\long\def\hom{\hat{\omega}}%

\global\long\def\hOm{\hat{\Omega}}%

\global\long\def\hdgg#1{\hat{#1}^{\dagger}}%

\global\long\def\cjg#1{#1^{*}}%

\global\long\def\hsgx{\hat{\sigma}_{x}}%

\global\long\def\hsgy{\hat{\sigma}_{y}}%

\global\long\def\hsgz{\hat{\sigma}_{z}}%

\global\long\def\hsgp{\hat{\sigma}_{+}}%

\global\long\def\hsgm{\hat{\sigma}_{-}}%

\global\long\def\hsgpm{\hat{\sigma}_{\pm}}%

\global\long\def\hsgmp{\hat{\sigma}_{\mp}}%

\global\long\def\dert#1{\frac{d}{dt}#1}%

\global\long\def\dertt#1{\frac{d#1}{dt}}%

\global\long\def\Tr{\text{Tr}}%